\PassOptionsToPackage{table}{xcolor}
\documentclass[biblatex]{sn-jnl}

\usepackage{graphicx}%
\usepackage{multirow}%
\usepackage{amsmath,amssymb,amsfonts}%
\usepackage{amsthm}%
\usepackage{mathrsfs}%
\usepackage[title]{appendix}%
\usepackage{xcolor}%
\usepackage{textcomp}%
\usepackage{manyfoot}%
\usepackage{booktabs}%
\usepackage{algorithm}%
\usepackage{algorithmicx}%
\usepackage{algpseudocode}%
\usepackage{listings}%

\usepackage[natbibapa]{apacite}

\usepackage{orcidlink}
\usepackage{placeins}
\usepackage[table]{xcolor} 
\usepackage{hyperref} 

\theoremstyle{thmstyleone}%
\begin{document}

\title[Article Title]{Conspiracy to Commit: Information Pollution, Artificial Intelligence, and Real-World Hate Crime}
\author*[1]{\fnm{Alberto} \sur{Aziani} \orcidlink{0000-0002-4745-7337}}\email{alberto.aziani@unimib.it}

\author[2]{\fnm{Michael V.} \sur{Lo Giudice} \orcidlink{0009-0008-5853-6058}}

\author[2, 3]{\fnm{Ali} \sur{Shadman Yazdi} \orcidlink{0009-0002-8317-8332}}

\affil[1] {\orgdiv{ Department of Sociology and Social Research}, \orgname{University of Milano - Bicocca}, \orgaddress{\city{Milan}, \country{Italy}}}

\affil[2] {\orgdiv{ Transcrime and Faculty of Social and Political Science}, \orgname{Università Cattolica del Sacro Cuore}, \orgaddress{\city{Milan}, \country{Italy}}}

\affil[3] {\orgdiv{ Department of Electronics, Information and Bioengineering}, \orgname{Politecnico di Milano}, \orgaddress{\city{Milan}, \country{Italy}}}

\maketitle

\makeatletter
\def\ps@myheadings{%
  \let\@oddfoot\@empty
  \let\@evenfoot\@empty
  \def\@oddhead{%
    \hfill
    \parbox[t]{0.9\textwidth}{\centering\footnotesize{%
      This is a pre-print version of the paper entitled \textit{Conspiracy to Commit: Information Pollution, Artificial Intelligence, and Real-World Hate Crime.} The published version is available at: \href{https://doi.org/10.1007/s10610-025-09629-w}{doi.org/10.1007/s10610-025-09629-w}%
    }}
    \hfill
  }%
  \def\@evenhead{\@oddhead}
}
\makeatother

\pagestyle{myheadings}

\pagestyle{myheadings}

\begin{abstract}

Is demand for conspiracy theories online linked to real-world hate crimes? By analyzing online search trends for 36 racially- and politically-charged conspiracy theories in Michigan (2015–2019), we employ a one-dimensional convolutional neural network (1D-CNN) to predict hate crime occurrences offline. A subset of theories—including the \textit{Rothschilds family}, \textit{Q-Anon}, and \textit{The Great Replacement}—improves prediction accuracy, with effects emerging two to three weeks after fluctuations in searches. However, most theories showed no clear connection to offline hate crimes. Aligning with neutralization and differential association theories, our findings empirically link specific racially-charged conspiracy theories to real-world violence. Just as well, this study underscores the potential for machine learning to be used in identifying harmful online patterns and advancing social science research. 

\medskip
\noindent \textbf{Keywords:} Information Pollution, Conspiracy Theories, Hate Crime, Deep Learning, Artificial Intelligence
\medskip

\noindent \textbf{Funding:} The research activities of certain authors for this paper received funding from the European Union's Horizon Europe research and innovation programme through the Fake News Risk MItigator (\href{https://fighting-fake-news.eu/}{FERMI}) project under grant agreement No. 101073980.
\end{abstract}



\section{Introduction}\label{Intro}

Conspiracy theories are defined as the allegation of a hidden plot between two or more powerful and malevolent actors that explains real social or political events \cite{douglas_understanding_2019, goertzel}, they exist on a spectrum, ranging from the highly implausible, like claims of a reptilian elite running the world, to the more believable, such as suspicions of specific corruption within government agencies. Despite their varying levels of plausibility, conspiracy theories generally allege shadowy motives by powerful groups. These groups can be governments (e.g., claims of the US having prior knowledge of the 9/11 attacks), large corporations (e.g., pharmaceutical companies suppressing definitive cures), ideological groups (e.g., a political party rigging elections), or even ethnic/religious minorities (e.g., Jewish dynasties controlling the global financial system) \cite{pantazi2022power}.

Conspiracy theories have been part of human society for centuries, often emerging in response to significant social or political events \cite{butter2014plots, van2017conspiracy}. Indeed, these theories can provide individuals with a framework to understand complex realities by attributing hidden motives to powerful actors, simplifying the intricacies of social and political events \cite{douglas, sunstein2009conspiracy}. However, the landscape of conspiracy theories has shifted with the advent of the internet and social media.

The internet has provided conspiracy theories, one of the oldest forms of information pollution, with new echo-chambers where their intricate rationalizations of events can go unimpeded by counterpoints \cite{cinelli}. Even when content moderation systems are present, the initial spread of false content is often unfettered \cite{miro-llinares_hate_2018} and evidence suggests that internet-driven radicalization may require rather short windows of exposure \cite{argentino}. That being said, empirical studies have so far only marginally approached this subject.

One of the many forms of information pollution that characterizes contemporary digital communication, conspiracy theories are especially troublesome \cite{cinelli}. Their reliance on unverifiable hypotheses is justified by the secrecy of the conspirators and, as such, allows the believer to cognitively overcome the lack of concrete evidence for, or even the outright impossibility of proving, the false information within a given theory \cite{douglas_understanding_2019, sunstein}. Although sometimes considered harmless, many conspiracy theories are discriminatory, painting segments of the population as devious actors, often based on race, religion, or social class \cite{lantian2018stigmatized}. In certain cases, the language used goes well beyond discrimination, actively employing hate-speech and encouraging believers to act \cite{baider2023covert}. Research has, in fact, identified that the proliferation of conspiracy theories poses a threat to society \cite{douglas2018conspiracy}. 

They deepen existing divides and erode trust in institutions, destabilizing social harmony and weakening democracy. Their reach extends even into the sphere of public health and economic well-being \cite{albertson2020conspiracy}. Exposure to conspiracy theories can, indeed, lead to \textit{epistemic lassitude}, a state of apathy towards seeking truth \cite{pennycook, sutton} which, in turn, fosters feelings of isolation and societal alienation \cite{goertzel}. By dismantling trust and fueling prejudice, conspiracy theories have the potential to weaken a society's social fabric and even foster collective violence \cite{douglas_understanding_2019}. 

In the past years, there has been growing anecdotal evidence behind the argument that conspiracy theories are a driver for criminal behavior. Despite sitting on the fringes of America's political landscape, an online movement referred to as \textit{Q-Anon} has, on at least five documented occasions, led to the commission of premeditated murder and domestic terrorism \cite{argentino}. Although its claims are demonstrably false, many consider \textit{Q} a vital informant, a supposed ex-member of Trump's first administration \textit{Q} reveals to the masses the terrible conspiracies afoot, usually involving Democrats and Hollywood elites. Likewise, \textit{the Great Replacement}, a theory originally spread by neo-nazis in Austria, has been growing in popularity across the West; ex-Fox News host, Tucker Carlson, mentioned the theory more than 400 times on-air, gifting it a new found legitimacy. In fact, it was cited in the manifesto of the Buffalo shooter, the man who intentionally targeted and killed 10 African Americans in a New York supermarket \cite{alrasheed2023cultural, Rose}. 

But can this relationship between conspiracy theories and crime be established empirically? And importantly, does it go beyond such infrequent, albeit dramatic, events; what is the impact the spread of conspiracy theories has on the crimes that occur in our cities daily? Is there a connection between the fear and anger fueled by these online narratives and real-life assaults, violence, and beatings?

Past work has shown that this link exists when placing online hate-speech and offline hate crimes under the \textit{academic microscope} \cite[e.g.,][]{burnap, muller_hashtag_2019, williams_hate_2019, muller_fanning_2021}. However, hate-speech and conspiracy theories differ. Although the latter may sometimes contain the former, hate-speech directly attacks specific groups with derogatory language and aims to incite violence or discrimination against them. At the core of conspiracy theories, instead, are unverified explanations for events, within a narrative structure, and the provision of alternative perspectives, potentially fueled by distrust in authority. Moreover, while online hate-speech appears to interact with real-world violence, causality between its diffusion in virtual environments and offline crimes is not always clear.

That being said, the influence of conspiracy theories on offline crimes remains understudied. Building on the work of \cite{LoGiudice}, this paper seeks to address this research gap, looking at the period beginning in January 01, 2015 and ending December 31, 2019, we leverage a set of deep learning algorithms to identify pathways connecting online demand for conspiracy theories and the commission of offline criminal activity. Utilizing Google Trends, we capture the online demand for racially- and politically-charged conspiracy theories in Michigan, US. Where \cite{LoGiudice} considered four theories, the investigation undertaken here expands to 36, a comprehensive picture of the contemporary conspiracy landscape. By analyzing the movement of these trends alongside reported hate crimes, we aim to explore the digital-to-offline reach of conspiracy theories.

This study offers five main contributions to the current understanding of conspiracy theories and hate crimes. First, it broadens the scope of conspiracy theory analysis. Rather than the traditionally examined areas of political systems, social cohesion, and health care \cite[e.g.,][]{jolley_pylons_2020, sutton, krouwel_does_2017, wood}, we shift the focus towards criminal behaviors.

Second, we advance the existing literature through the inclusion of real actions, as opposed to merely the intention to act. Previous research has primarily examined the potential for conspiracy theories to influence individuals' beliefs and intentions, or how exposure to conspiracy theories shapes trust in government and support for anti-establishment movements \cite{douglas, sutton}, leaving a gap in understanding how these sentiments translate into realized actions. By analyzing reported hate crimes, our study moves beyond self-reported intentions or perceptions to reality, examining real-world behaviors and their connection with the demand for conspiracy theories online, captured through search trends.

Just as well, the body of current literature is at a micro-level, that is individual-level, of analysis \cite[e.g.,][]{douglas_are_2021, rottweiler_conspiracy_2022}. We alternatively adopt a macro-approach, examining the potential harm related to online conspiracies at a societal-level. By analyzing data at an aggregated level, statewide, we complement existing studies, providing both a quantitative and macro reinforcement for scholars micro, survey-based findings.

Fourth, existing research on the topic of information pollution, a phenomenon that encompasses conspiracy theories, tends to focus on the supply of disinformation (i.e., fake news articles, social media posts, etc.) and how people perceive them. Our work examines the demand side, specifically the actual seeking out of these theories, a factor found particularly relevant for Islamist radicalization via the internet \cite{frissen}.

Furthermore, our work contributes to the growing trend of applying machine learning to crime-related issues \cite[e.g.,][]{aziani2023convergence, campedelli2022explainable, gyourko2024predictors, wang2023pursuit, wheeler2021mapping, daniele, evangelatos} and represents one of the first attempts in the social sciences to isolate the contribution of features to the prediction capacity of a model as a signal for the presence of a relationship \cite[see][]{LoGiudice, castro}.

This paper is organized as follows; the \nameref{LitRev} covers key academic insights into conspiracy theories and the relationship between online content and offline crimes, leading us to our central research question: \textit{is there a relationship between searching online for conspiracy theories and offline hate crimes}? This section also introduces our research hypothesis and the theoretical reasoning supporting it. The \nameref{sec3} section details the deep learning algorithms we applied, our strategy for investigating the relation between conspiracies and hate crimes, and the publicly available data we utilize. \nameref{sec4} presents the primary findings of our analysis: while the majority of conspiracy theories investigated show no significant relationship with offline hate crimes, there exists a subset whose trends enhance the predictive accuracy of our model. In the \nameref{sec5}, we discuss these findings and their meaning in the context of the existing literature, highlight the limitations of our study, and suggest directions for future studies. The concluding section offers a summary of our research and outline a policy implication of our finding.

\section{Literature Review}\label{LitRev}

\subsection{Prevalence and Impact of Conspiracy Theories}

Conspiracy theories are pervasive across societies. Survey data reveal that 60\% of Americans believe that the Central Intelligence Agency was involved in President Kennedy's assassination \cite{enders}. More recent conspiracy theories, such as the claim that 5G infrastructure causes disease (including COVID-19), further illustrate their widespread appeal \cite{jolley_pylons_2020, yang2021all}. These theories often gain traction due to their ability to provide alternative narratives in times of uncertainty \cite{douglas_understanding_2019, butter2020routledge}.

Research shows a link between belief in conspiracy theories and psychological factors such as anomie and disillusionment. For example, people who accept conspiracy theories exhibit higher levels of societal distrust and greater acceptance of political violence \cite{krouwel_does_2017, imhoff_2020}. This psychological vulnerability might create fertile ground for exploitation, particularly by extremist groups, which can use conspiracy theories to reinforce ideological views and justify actions as neutralization techniques \cite[see][]{sykes1957techniques}. As evidenced by the far-right's promotion of the \textit{Kalergi Plan} and the \textit{Q-Anon} movement \cite{argentino, quassoli2024risk, rottweiler_conspiracy_2022}. Moreover, conspiracy beliefs are associated with discriminatory attitudes and behaviors, particularly when targeting specific groups \cite{bilewicz_harmful_2013, jolley_exposure_2020}. A meta-analysis showed that personality traits like narcissism and social identity motives significantly correlate with belief in conspiracies, suggesting a deeper psychological underpinning to their adoption \cite{goertzel}.

\subsection{Link Between Online and Offline Behaviors}

The Internet has amplified the reach and influence of conspiracy theories. Studies suggest that online engagement with extremist content, including conspiracies, can predict radicalization better than traditional factors such as juvenile delinquency or moral disengagement \cite{frissen}. Similarly, in terms of the access of information online leading to offline actions, \cite{Uscinski_2018} notes that consuming a conspiracy theory often guides to its adoption.

However, while online platforms have mechanisms such as content moderation to limit the spread of false information, these efforts often fail to prevent the rapid and widespread dissemination of conspiracy theories, particularly in their early stages \cite{miro-llinares_hate_2018}. This initial surge enables conspiracy theories to reach a large audience before interventions take effect, solidifying and strengthening pre-existing beliefs \cite{cinelli, wood}. Recommendation algorithms play a role in this by prioritizing engagement metrics that expose users to extremist content, including conspiracy theories \cite{shin2024misinformation}. By facilitating this unrestrained dissemination, the Internet functions as a \textit{radicalizing multiplier}, not only amplifying conspiracy narratives, but also eroding individuals’ connections to civil society and weakening barriers to extremist actions \cite{lee, ricci2024surveying, rottweiler_conspiracy_2022}.

Empirical evidence linking conspiracy theories to offline crimes remains limited \cite[see][]{miro-llinares_misinformation_2023}. However, studies examining the role of online hate-speech in offline violence—such as the correlation between racially charged posts on X (formerly Twitter) and hate crimes \cite{williams_cyberhate_2016, williams_hate_2019}—suggest similar mechanisms may be at play with conspiracy theories. For instance, anti-refugee sentiment on Facebook has been shown to predict crimes against refugees in Germany, highlighting the connection between online rhetoric and real-world consequences \cite{muller_fanning_2021}. Although similar research on conspiracy theories is still scarce, these findings concerning hate-speech suggest that conspiracy theories may likewise influence offline behavior through similar mechanisms.

Even though conspiracy theories are not inherently synonymous with hate-speech, their narratives might overlap. Consider theories positing secret cabals controlling governments, they may not directly incite hatred but they do foster environments of fear and distrust, which can lead to violence \cite{douglas_are_2021, jolley_exposure_2020}. This overlap is particularly evident in extremist propaganda, where conspiracies are used to justify hostility toward marginalized groups.

Conspiracy theories undermine trust in authority and deepen societal divisions; however, it remains unclear how the demand for conspiracy-related online content correlates with offline hate crimes. Current studies rely primarily on self-reported data, which is prone to expressive responses and inaccuracies. For example, respondents may intentionally misrepresent their beliefs to align with perceived norms or biases \cite{douglas_understanding_2019}. The methodological limitations in previous research call for an approach that focuses on realized actions rather than self-reported attitudes.

\subsection{Research Gap and Hypothesis}

While there is a growing body of literature addressing the psychological underpinnings and societal impacts of conspiracy theories, particularly during events such as the COVID-19 pandemic and the surge of 5G-related conspiracies, a notable gap persists in empirical research examining the direct relationship between conspiracy theories and offline crimes. Most existing studies face challenges in linking online behaviors to real-world actions, often relying on self-reported attitudes toward conspiracies rather than direct behavioral data. This reliance is problematic, as individuals can misrepresent their beliefs in politically-charged contexts, raising concerns about the accuracy and reliability of these studies.

In response to calls from previous research, this study seeks to address this gap by investigating the relationship between the demand for conspiracy theories online and the occurrence of hate crimes offline. Given the established connections between conspiracy beliefs and feelings of anomie or disillusionment, as well as observed correlations between online hate-speech and offline violence, we hypothesize that incorporating data on online searches for conspiracy theories will enhance the predictive accuracy of deep learning models in forecasting hate crime occurrences.

Specifically, we propose that trends in online searches for conspiracy theories will serve as valuable predictors, improving the performance of models that would otherwise exclude this data.  We expect that online demand trends for conspiracy theories (captured through online searches) will significantly enhance the ability of deep learning algorithms to predict hate crimes offline compared to models without this information. To reiterate, the following behavior is expected: \textit{the trends in online searches for conspiracy theories will improve the deep learning model’s ability to predict hate crimes offline}. An eventual increase in predictive performance will be an indication of a relationship between the demand for conspiracies and offline hate crimes.

Our expectations align with the \textit{neutralization theory} by \cite{sykes1957techniques}, which posits that individuals rationalize or neutralize their deviant behavior by appealing to justifications that allow them to overcome societal norms. Conspiracy theories, often framed in ways that delegitimize targeted groups or institutions, may provide such rationalizations by portraying targets as threats or villains within a narrative structure. Narratives that could empower individuals predisposed to prejudice or bias to commit hate crimes while maintaining a moral self-image.

Furthermore, Sutherland's (\cite{sutherland1947principles}) \textit{differential association theory} offers a complementary perspective. This theory suggests that criminal behavior is learned through interaction with others, particularly through the communication of values, attitudes, and rationalizations favorable to crime. Conspiracy theories, which often circulate within ideologically homogeneous online and offline communities, may serve as mechanisms of such transmission, fostering an environment in which hate crimes are perceived as acceptable or justified responses to the alleged threats posed by out-group members.

\section{Methodology}\label{sec3}

Our study investigates whether online searches for conspiracy theories predicts offline hate crimes by analyzing temporal relationships between 36 conspiracy-related Google search trends and weekly hate crime data in Michigan (2015–2019). A one dimensional convolutional neural network (1D-CNN) model is selected for its ability to detect localized temporal patterns, trained to predict hate crimes using historical crime data, seasonal controls, and conspiracy search trends. To isolate the predictive value of conspiracy-related searches, models incorporating these trends are compared to a baseline model trained only on historical crime data and seasonality. Improved accuracy (reduced prediction error) when conspiracy trends are included suggests a statistical association. To ensure results reflect genuine temporal relationships—not spurious correlations—a feature permutation test disrupts the time-order of conspiracy search data while retaining numerical properties. If model accuracy degrades when trained on shuffled data, it confirms that predictive power depends on the genuine temporal sequence of searches, not static numerical artifacts.

\subsection{Data Collection and Pre-Processing}

Given our objective, to investigate whether there is a link between the demand for conspiracy theories online and registered hate crimes offline, we compile a list of politically- and racially-charged conspiracies (and conspiracy-related terms) from the \href{https://extremismterms.adl.org}{Anti-Defamation League’s Glossary of Extremism and Hate}.\footnote{These theories/theory-related terms are: Adrenochrome; Deep State; Deep Underground Military Bases; D.U.M.B.s (the acronym for \textit{deep underground military bases}); Died Suddenly; Event 201; Fall Of The Cabal; Frazzledrip; George Soros; JQ (meaning \textit{the Jewish question}); Kalergi Plan; Kristallnacht; Mole Children; Obama Kenya; Pedogate; Pizzagate; Q Sent Me; Q-Anon; RAHOWA (meaning \textit{racial holy war}); Rothschild Family (searched as \textit{Rothschilds}); Seth Rich Murder; Takiya; Ten Days Of Darkness; The 14 Words; The Cabal; The Deep State; The Great Awakening; The Great Replacement; The Great Reset; Trump 19th President; Tunnel Children; Tuskegee Syphilis Study; U.S.S. Liberty; We Are The Storm; White Genocide; Zionism.} For each conspiracy, we define search terms determined to be most closely related to the conspiracies. If the conspiracy has a clear endogenously given name, the search term corresponds to it (e.g., \textit{the Great Replacement}). In cases where the theory has no endogenously prescribed name, the search term chosen best matches the key words of the theory. For example, the conspiracy theory commonly referred to as \textit{the birther movement}, has a name prescribed to it by American media outlets and politicians, as opposed to the followers of the theory (i.e., an exogenous name). In this case, we opt to use the key words \textit{Obama Kenya}, since former President Obama is the subject of the theory and Kenya a central element in its content (the former president's theorized place of birth). There is also the inclusion of terms related to real events or persons (i.e., not conspiracies) that according to the Glossary of Extremism and Hate were relevant to modern conspiracy/extremist discourse; the limitations of such terms being included is discussed at detail in the subsection discussing limitations.

The online demand for each conspiracy is then collected from Google Trends, a publicly accessible tool offered by Google that allows one to see the popularity of specific terms searched on Google in a given period of time. Given Google’s dominant market share among search engines, the trends in this search engine are representative of the interest of the online public in the topic. In the US, the average market share for Google's search engine during our period of interest (i.e., 2015-2019) was 85.9\%, effectively granting Google a monopoly in the market \cite{AntitrustDivision2021}. Consequently, scholars extensively use Google Trends for predictive purposes, ranging from electoral results to disease outbreaks, products' market share, and financial investments, among others \cite[e.g.,][]{ahmed2017financial, behnert2024can, bong2020analysis, prado2021google, mager2023european, morsy2018prediction}. 

All search trends are collected from January 1$^{\text{st}}$, 2015 until December 31$^{\text{st}}$, 2019 at weekly intervals and geo-located to the state of Michigan via IP addresses. The period selection has a three-fold motivation. Firstly, the COVID-19 lockdown orders, issued in early 2020, had an inflationary effect on online traffic \cite{masaeli}. At the same time, lockdown orders prevented people from leaving their home, which affected the frequency of crimes \cite{boman, campedelli}. Thirdly, Google Trends only provides weekly observations for a selection four-years or less, with larger samples, it aggregates the data to a monthly-level. Hence, 2015 - 2019 is chosen, the longest and most recent sample of weekly trends possible while ensuring that COVID-19 and its exceptional circumstances are excluded. 

Michigan is selected due to the comprehensive \href{https://cde.ucr.cjis.gov/LATEST/webapp/#/pages/explorer/crime/hate-crime}{dataset} made openly available by the Federal Bureau of Investigation, wherein hate crime occurrences are reported at the incident level and annotated with a date (Fig. \ref{fig1}). Several other states were considered, in some cases the date was not provided for a significant number of observations; while for others, the number of searchable theories (via Google Trends) was insufficient and would limit the scope of the investigation. Therefore, for the present study, Michigan proved to be the most suitable in terms of crime data quality and quantity of Google Trend results. 

The hate crime data are reported criminal acts, varying in type, that were, when prosecuted, tagged as having been carried out as an act against the victim or their property due to some identifiable characteristic of the victim (e.g., race, religion or ethnicity). Hate crimes in Michigan fall under Michigan Penal Code, Act 328 of 1931, § 750.147b pertaining to ethic intimidation. It is important to note that while the law is specific to ethnic intimidation, it has, in practice, been used for prejudices other than against one's ethnicity. During the four-year period sampled, there is a registered hate crime in all 262 weeks examined (Table \ref{tab1}).

\begin{figure}
    \centering
    \includegraphics[width=.7\linewidth]{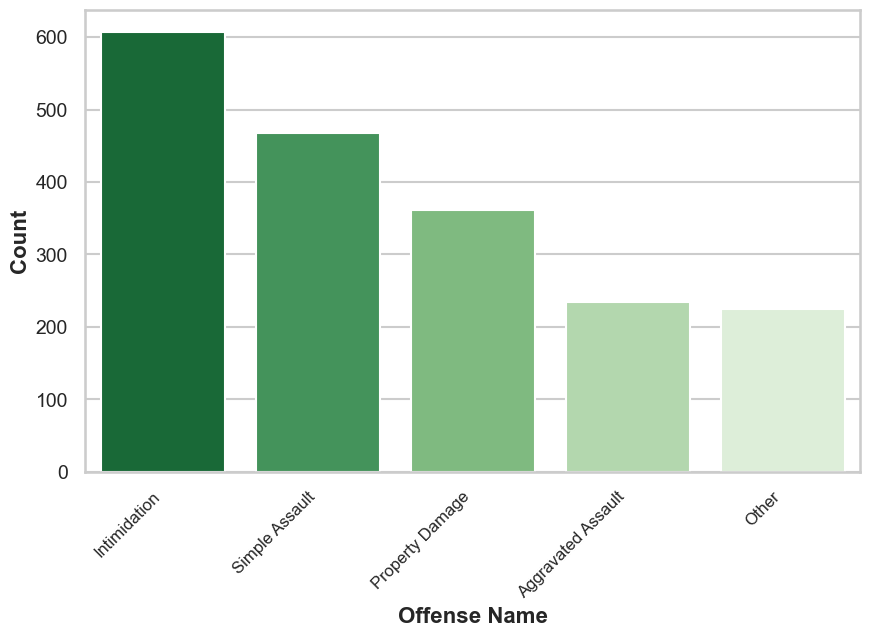}
    \caption{\textbf{Crime types associated with hate crime occurrences in Michigan from 2015--2019.}}
    \label{types}
\end{figure}

\begin{figure}
    \centering
    \includegraphics[width=.7\linewidth]{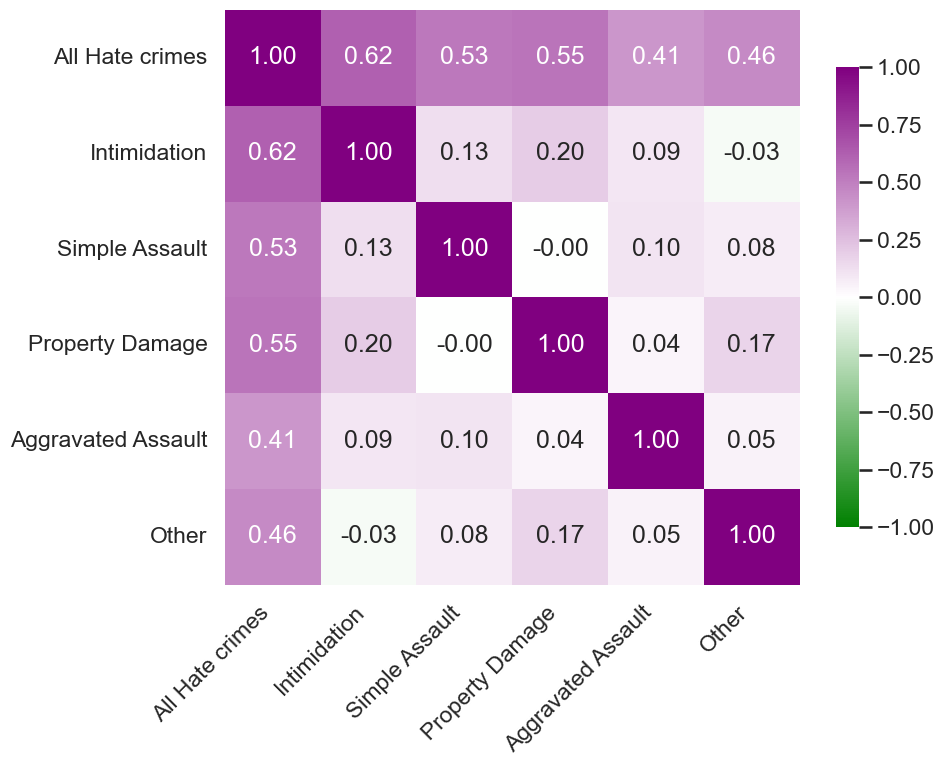}
    \caption{\textbf{Correlation of different hate crime types with the aggregated ones.}}
    \label{crimeCorr}
\end{figure}

The original hate crime data categorizes offenses by crime type and sometimes by bias type (e.g., targeting race or sexual orientation), with some cases featuring multiple crime times and/or a plurality of victims/offenders. As is typical for American incident-level crime data, the 'incident' may include a plurality of criminal offences. For this study, we analyze hate crimes collectively without distinguishing by crime or bias type. While labeling biases could allow to align specific hate crimes with relevant theories (e.g., antisemitic crimes with antisemitic conspiracy theories), we opt against this approach. Previous research indicates that belief in a discriminatory theory targeting one group often predicts prejudices against other groups as well \cite{jolley_exposure_2020}. This suggests conspiracy theories may broadly harm inter-group relations and fuel social instability \cite{douglas_are_2021, sutton}. To avoid overlooking this wider societal impact, we aggregate all hate crimes, regardless of bias type, in our analysis. Moreover, a review of the crime types present in the data provides confidence that our approach of taking all hate crimes does not pose a risk for the significant inclusion of irrelevant crime types. Intimidation, assault (simple and aggravated) and property damage comprise the vast majority of observations (Fig. \ref{types} and Fig. \ref{crimeCorr}).

\begin{figure}[h]
 \centering
 \includegraphics[width=1\linewidth]{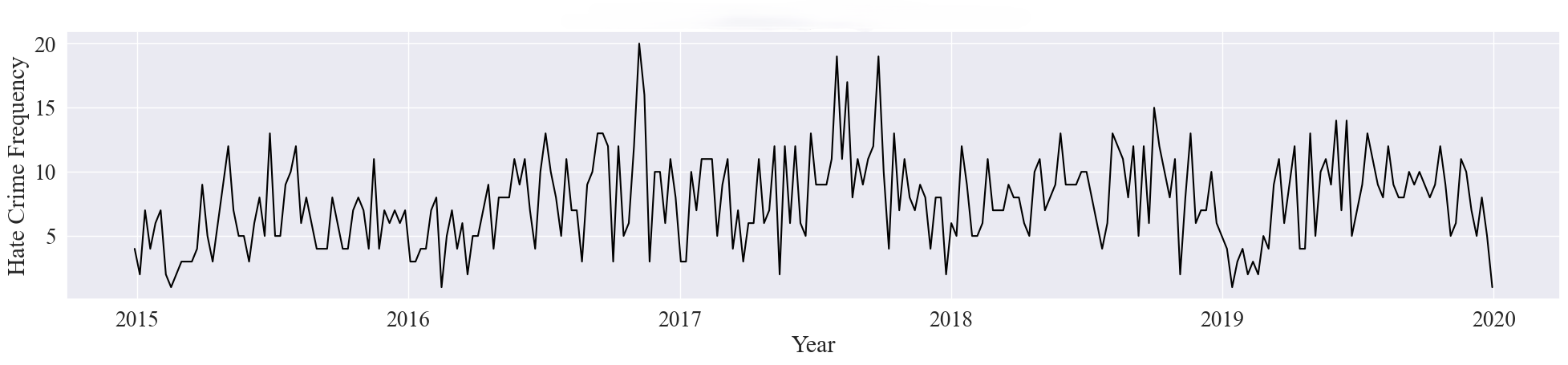}
 \caption{\textbf{Weekly Hate Crime Frequency in Michigan, 2015--2019}.
 Weekly reported hate crime occurrences in Michigan from 2015 to 2019, illustrating temporal trends in hate crime frequency during the study period.}
 \label{fig1}
\end{figure}

\begin{table}[h]
 \centering
 \caption{\textbf{Weekly Hate Crime Statistics in Michigan, 2015--2019}}
 \begin{tabular}{ccccccccc}
 \toprule
 {} & Number of Weeks & Average & Std & Min & Max \\
 \midrule
 \textbf{Weekly Hate Crimes} & 262 & 7.67 & 3.45 & 1 & 20 \\
 \midrule
 \end{tabular}
 \label{tab1}
 \footnotetext{The descriptive statistics for registered hate crimes in Michigan, every seven days, beginning January 1, 2015 and ending December 31, 2019. There was no seven-day period during the four years sampled without an observed, hate crime.}
\end{table}

The crimes are aggregated to weekly totals to ensure both the hate crime and conspiracy time-series have the same structure. Then, both time-series undergo scaling to ensure consistency in their value ranges. Thus, with our time periods, \textit{t}, set to weeks (every seven days beginning on 01/01/2015), we have a \textit{T} = 262, with each \textit{t} having a value for the online demand of each conspiracy (as provided by Google via the pytrends API) and for the number of hate crimes between 0 and 100 (scaled relative to the global maximum of hate crime data). We also add a set of dummy variables, noting the week and month of the year. These act as month and week controls and account for seasonal fluctuations in both online searches and hate crimes \cite[see][]{andresen, mcdowall, torres, freeman}.

The data is then divided into training and validation sets, allocating 80\% for training and 20\% for validation, a standard division to train a deep learning model. Additionally, a windowing approach is employed, segmenting five weeks of hate crime data, along with their corresponding conspiracy trends, as the models' input \cite{torres, freeman}. Splitting the data into training and validation sets ensures that a model has not studied the data on which its accuracy is being tested, and the windowing technique was applied independently to both sets to further ensure there is no overlap. The models then generate hate crime occurrence predictions for the next four weeks within the input window. We chose a four-week time span to maintain a balance between our capacity to capture medium-term effects and the models' complexity.

\subsection{Model Selection}
\FloatBarrier
\begin{figure}
 \centering
 \includegraphics[width=0.8\linewidth]{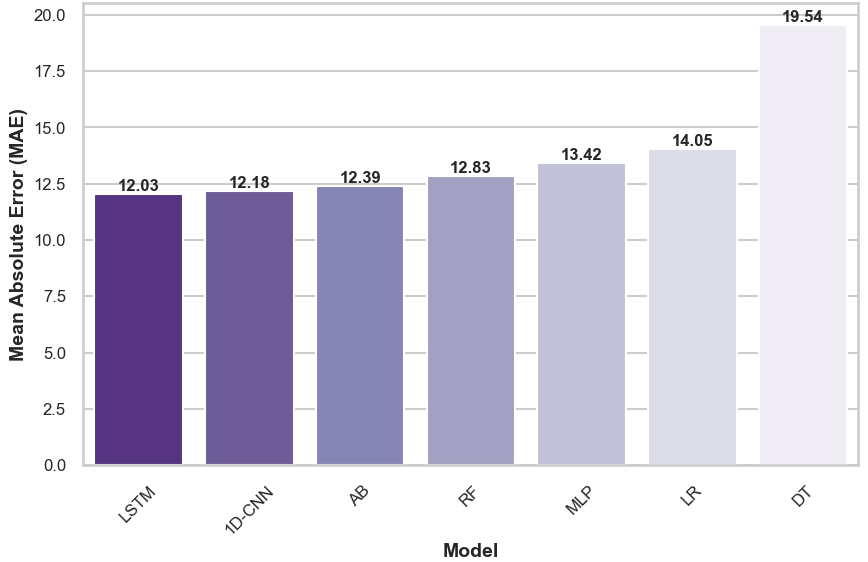}
 \caption{\textbf{Model Performance (scaled MAE) for Test Set of Hate Crimes}.
 Comparison of machine learning and deep learning models' performance on the hate crime test set, evaluated using scaled mean absolute error (MAE).}
 \label{fig_model_sel}
\end{figure}

In determining the most suitable model to use as a baseline, we trial a range of machine learning and deep learning models to identify which one achieves the best predictive performance before introducing online demand for conspiracies and testing our hypothesis that searches for conspiracies informs hate-crime prediction. The evaluated model architectures are the following: long-short term memory (LSTM) \cite{hochreiter1997long}, 1D-CNN \cite{kiranyaz2015real}, AdaBoost (AB) \cite{freund1997decision}, random forest (RF) \cite{breiman2001random}, multi-layer perceptron (MLP)  \cite{rosenblatt1958perceptron}, linear regressor (LR) \cite{montgomery2021introduction}, and decision tree (DT) \cite{quinlan1986induction}. For the machine learning models, we utilize the MultiOutputRegressor wrapper to handle multi-output predictions. Each model is trained and tested on the dataset, and their performance is assessed based on their scaled mean absolute error (MAE). 

1D-CNN (MAE of 12.18) and LSTM (MAE of 12.03) are the best performing among the tested models (see Fig. \ref{fig_model_sel}), in line with our expectations as said architectures are known for a superior capacity in capturing temporal dependencies and patterns in data \cite{bai2018empirical, santry, wang2017permutation}. Although LSTM achieves the lowest MAE, we observe that 1D-CNN is better able to understand and predict localized peaks in hate crime occurrences; as noted in previous research \cite[see][]{LSTM}. A better performance in appreciating local peaks is significant given the context of our study, as previous studies suggest that only a brief period of exposure, online, is needed before individuals act offline \cite{argentino}. As such, we chose to subject LSTM and 1D-CNN to an additional level of evaluation, employing a structural similarity index measure (SSIM). 

SSIM is a correlation-based metric that assesses both the location and intensity of local peaks by comparing the evolutionary structure of ground truth values with predictions. While traditionally applied to image analysis, studies have demonstrated its effectiveness for evaluating similarities in time-series data \cite{SSIM}. For our purposes, it serves to provide an understanding of whether the 1D-CNN or the LSTM better capture local peaks. SSIM scores ranges from -1 to 1, with 1 indicating a perfect match between ground truth and prediction. The results reveal a significantly higher SSIM score for the 1D-CNN compared to the LSTM, thereby affirming the superior performance of the former. The forecasting task is instrumental: we need a stable learner whose error
surface is sensitive to informative inputs, not the absolute lowest possible
MAE. Based on this comparison, the 1D-CNN model is chosen for our analysis. 

\begin{figure}
 \centering
 \includegraphics[width=\linewidth]{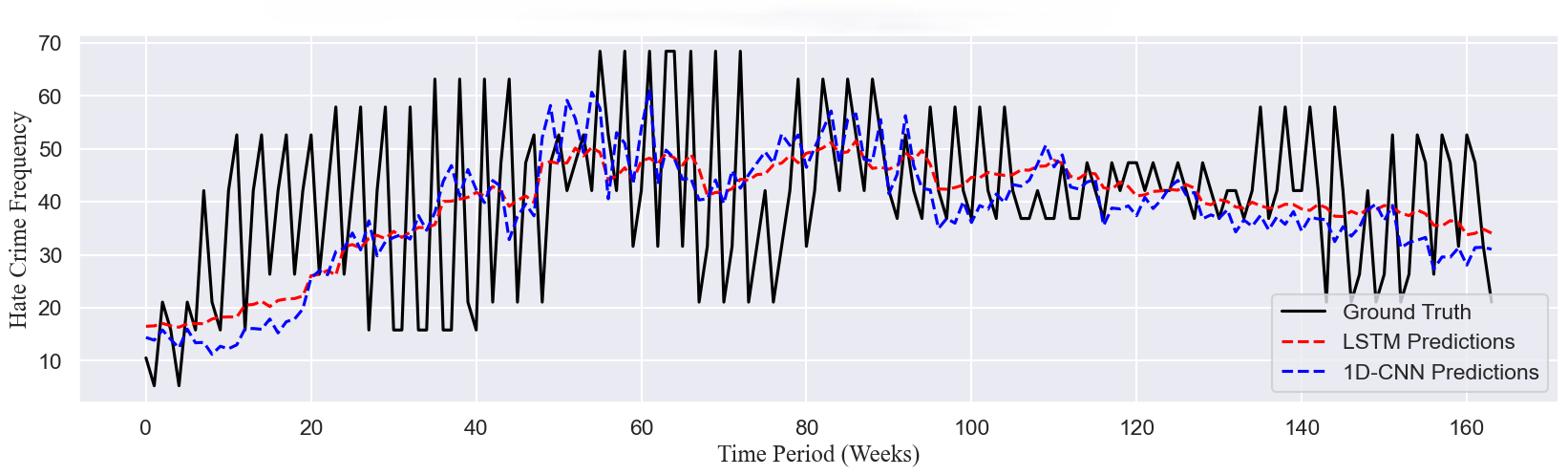}
 \caption{\textbf{LSTM and 1D-CNN Predictions Compared with Ground Truth Values (Test Set)}.
 Comparison of LSTM and 1D-CNN model predictions against ground truth hate crime occurrences in the test set.}
 \label{fig3}
\end{figure}

\begin{figure}[h]
 \centering
 \includegraphics[width=1\linewidth]{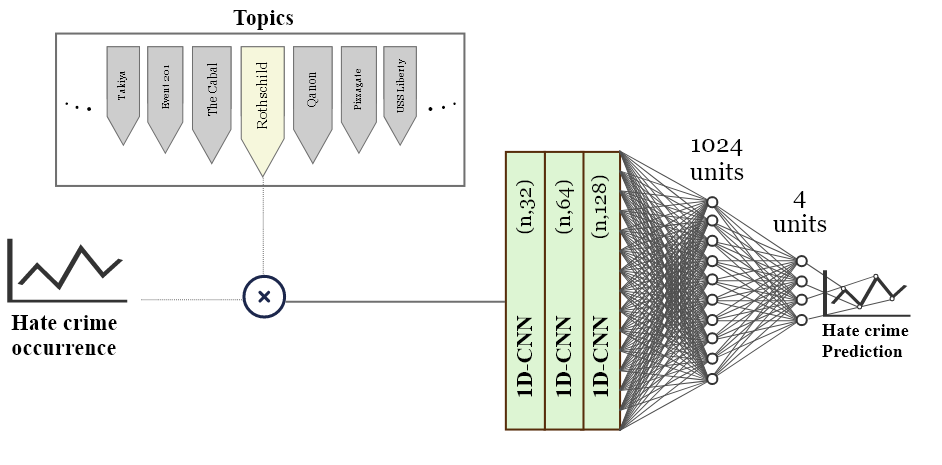}
 \caption{\textbf{Visualisation of 1D-CNN Model Architecture and Pipeline}.
 The selected 1D-CNN model consists of three convolutional layers with increasing filter sizes (32, 64, and 128) to progressively capture complex patterns in the input data. Each convolutional layer applies the ReLU activation function, while fully-connected layers (1024 and 4 units) refine the learned features for final hate crime prediction. No max-pooling layers are used to prevent excessive information loss. Training employs early stopping and learning rate reduction strategies to optimize convergence and prevent overfitting. Each model is trained separately per theory to assess its individual predictive contribution.}
 \label{fig2}
\end{figure}

The selected 1D-CNN architecture consists of three convolutional layers, each increasing in the number of filters: 32, 64, and 128. These filters can be conceptualized as small grids that progressively capture more complex features in the input data. Each convolutional layer employs the rectified linear unit (ReLU) activation function \cite{agarap}. Given the relatively small dataset, max-pooling layers were deliberately omitted to preserve the weekly
resolution of the input series; instead, model capacity is controlled with dropout ($p\!=\!0.30$) and early-stopping. This light-capacity design follows common small-sample time-series guidance \cite[e.g.,][]{bai2018empirical, ismail2019deep, ismail2020inceptiontime}. Instead, the model relies solely on convolutional transformations to extract meaningful patterns. As the network deepens, the increasing number of filters enhance its ability to learn a broader range of features and capture intricate patterns in the data \cite{goodfellow}. Proceedingly, the model incorporates two fully-connected layers; the first is composed of 1024 units, also utilizing the ReLU activation function, while the second serves as the output layer, comprising four units with a linear activation function (Fig. \ref{fig2}). The convolutional operation in the \begin{math}i^\text{th} \end{math} layer of the \begin{math}j^\text{th}\end{math} set can be represented as follows:
\begin{equation}\label{Equation 2}
 Z[i,j] = W[i,j] \times X + b[i,j]
\end{equation}
\begin{equation}\label{Equation 3}
 A[i,j] = \text{ReLU}(Z[i,j])
\end{equation}
Where \begin{math} Z[i,j] \end{math} is the convolutional output layer, a product of \begin{math}W[i,j]\end{math}, the weight matrix for the \begin{math}i^\text{th}\end{math} layer of the \begin{math}j^\text{th}\end{math} set, $X$, the input data, and \begin{math}b[i,j]\end{math}, the bias for the \begin{math}i^\text{th} \end{math} filter in the \begin{math}j^\text{th}\end{math} set. \begin{math}A[i,j]\end{math} is, thus, the transformed output, through the application of the ReLU activation.

The fully-connected layers can be represented as follows: 
\begin{equation}\label{Equation 4}
 Z[k] = W[k] \cdot A[k-1] + b[k]
\end{equation}
Here, \begin{math}Z[k]\end{math} is the output of the \begin{math}k^\text{th}\end{math} fully-connected layer, a product of \begin{math}W[k]\end{math}, the weight matrix for the \begin{math}k^\text{th}\end{math} fully-connected layer, \begin{math}A[k-1]\end{math}, the output of the previous layer, and \begin{math}b[k]\end{math}, the bias for the \begin{math}k^\text{th}\end{math} fully-connected layer.

During the training phase of the model, batch size of 4 was used with continuous training until convergence, monitored through early stopping. The early stopping mechanism was configured with a patience of 15 epochs, restoring the best model weights based on the validation loss. Additionally, a learning rate reduction on plateau strategy was applied, reducing the learning rate by a factor of 0.1 if validation loss did not improve for 5 consecutive epochs (with a minimum learning rate threshold set at 1e-9). The model was trained using mean squared error (MSE) loss and optimized with the Adam optimizer, ensuring stable convergence \cite{kingma}. The validation dataset was utilized to assess generalization performance and mitigate over-fitting. To further counteract over-fitting, early stopping and learning rate reduction techniques were implemented, as well as the incorporation of a dropout layer, between the fully-connected layers, to enhance model generalization \cite{prechelt, park}. Models undergo training and evaluation independently for each topic. To ensure consistent and reproducible results throughout the training process, the GlorotNormal initialization method is utilized with a fixed seed \cite{narkhede}. By initializing the model weights with a fixed seed, randomness during training is eliminated, ensuring that the observed results are solely influenced by variations in the input data. That is, the number of hate crimes in the period being used to test accuracy will not vary between models and is therefore not the explanation for any observed variations in accuracy.

\subsection{Analytical Strategy} \label{Feature Permutation Test}

The model selection procedure leads us to adopt 1D-CNN to explore the relationship between online demand for conspiracy theories and offline hate crimes. We train 36 separate 1D-CNN models-one for each conspiracy-where each model is provided the time-series of weekly search activity for a specific conspiracy theory, historical hate crime data, and seasonal controls (week/month dummy variables). Equation \ref{Equation 1} formalizes this relationship, where hate crimes at time $t$ ($H_t$) depend on three factors: historical hate crime patterns ($H_{0-t}$), the time-series of online search activity for a specific conspiracy ($C_{0-t}$), and seasonal effects ($S_t$):

\begin{equation}\label{Equation 1}
H_t = f(H_{0-t}, C_{0-t}, S_t)
\end{equation}

To evaluate the predictive value of conspiracy-related search trends, we compare the performances of models trained with conspiracy data against a baseline model trained exclusively on historical hate crime data and seasonal controls (week/month dummy variables) without conspiracy-search features. If incorporating conspiracy-related searches reduces the model’s prediction error (compared to the baseline), we infer that the conspiracies in question carry unique information about future hate crime occurrences. Reduced prediction errors imply that the model leverages the conspiracy's search trends to enhance accuracy, indicating a statistical association between the two variables \cite{castro, del2024robust, lundberg2017unified, molnar2020interpretable, scholkopf2021toward}. Each conspiracy-search trend is then tested at four different lags (-1, 1, 2, 3), to control for potential delayed or reverse relations between demand online and actions offline. 

Deep neural networks, including CNNs, can capture complex, non-linear patterns but often lack transparency regarding how they arrive at specific predictions. A key risk is that a model might exploit spurious correlations or purely numeric coincidences in the data rather than meaningful temporal signals. In other words, even if a model’s performance improves when provided an online demand trends for a conspiracy theory, it is not guaranteed that the improvement stems from the time-varying pattern of said trend (i.e., its shape over time); in principle, a highly flexible model could latch onto simple numeric properties. Therefore, to verify that improvements indeed arise from the temporal structure of conspiracy searches (and not merely from static numerical artifacts), we apply a feature permutation test \cite{breiman2001random,canovas2009permutations,molnar2020interpretable}.

The feature permutation test involves randomly shuffling the values of each conspiracy-theory time series while keeping all other features unchanged. By comparing the forecast error of the model trained with the original time-series to that of a model trained with the shuffled series, we determine whether the predictive advantage disappears when the temporal order is disrupted \cite{breiman2001random, canovas2009permutations}. If accuracy declines after permutation, it confirms that the original temporal structure, rather than any static numerical property, was responsible for the observed predictive power. To ensure robustness, each test is repeated three times per conspiracy theory. A genuine relationship between online searches and hate crimes is established only if the model consistently relies on the unaltered time series for improved accuracy, rather than producing similar results with shuffled versions.

In standard permutation importance, a single model is typically trained on the original data and then evaluated on permuted test data; however, in our time-series context we re‑train the model for each permutation to fully capture the impact of disrupted temporal ordering on both the learning and prediction processes. This re‑training approach ensures that any performance degradation directly reflects the loss of the feature’s predictive power, because the model is forced to learn from scratch without access to the intact temporal structure. In contrast, simply re‑evaluating a model trained on unpermuted data could leave residual effects from the original temporal dynamics, thereby underestimating the true importance of the feature.

In summary, for each conspiracy, a corresponding search term is identified. From which, the Google trend from 2015 until 2019 is collected. Five 1D-CNN models are trained, creating independent interpretations of the data sequences and any potential nexuses between them. The models varied only in the lag introduced to the conspiracy's trend, with each model featuring a lag between zero and three periods. The fifth model instead pushed the conspiracy trend forward (i.e., lag -1), to provide insight on the direction of the relationship.

The accuracy of these models, conceptualized as the scaled MAE for a prediction of unseen hate crime occurrences, is compared to a baseline model, not provided any conspiracies' trend. If the MAE of a model with a trend is lower than the baseline, it is considered as prediction improving and indicative of a relationship between the relevant conspiracy and hate crimes. To ensure the robustness of our findings, said trends are then subjected to a feature permutation test, as described above, with the ultimate goal of revealing the legitimacy of the feature's influence on model performance \cite{castro, konig}.

We can represent the feature permutation test as follows. Let \(\mathcal{D} = \{(X_t, y_t)\}_{t=1}^{T}\) be a time-series dataset, where:

\begin{equation}
    X_t = (x_{t1}, x_{t2}, \dots, x_{td})
\end{equation}

denotes the feature vector at time \(t\), with \(x_{tj}\) representing the \(j^\text{th}\) feature (e.g., search trend for conspiracy theory \(j\)), and \(y_t\) denoting the target variable (e.g., hate-crime count) at time \(t\). Let \(M\) be a trained forecasting model (e.g., 1D-CNN), producing predictions:

\begin{equation}
    \hat{y}_t = M(X_t)
\end{equation}

The MAE of \(M\) on \(\mathcal{D}\) is given by:

\begin{equation}
    \text{MAE}(M, \mathcal{D}) = \frac{1}{T} \sum_{t=1}^{T} |y_t - \hat{y}_t|
\end{equation}

For the baseline model, we compute the MAE on the original dataset as:

\begin{equation}
\text{MAE}_{\text{original}} = \text{MAE}(M, \mathcal{D})
\end{equation}

This value represents the performance of the model \(M\) trained on all features in the original dataset \(\mathcal{D}\) and does not depend on any search trend for conspiracy theory \(j\).

To evaluate the predictive contribution of each search trend for conspiracy theory \(j \in \{1, \dots, d\}\), the permutation process is repeated \(K\) times (e.g., \(K=3\)). In each iteration \(k\), a permuted dataset \(\mathcal{D}_{\text{perm}, j, k}\) is generated by shuffling the temporal values of the search trend for conspiracy theory \(j\):
\begin{equation}
\mathcal{D}_{\text{perm}, j, k} = \{(X_{t,j,k}, y_t)\}_{t=1}^{T}, \quad \text{where } X_{t,j,k} = (x_{t1}, \dots, x_{\pi_k(t)j}, \dots, x_{td})
\end{equation}
with \(\pi_k: \{1, \dots, T\} \to \{1, \dots, T\}\) being a random permutation of time indices. A new model \(M_{\text{perm}, j, k}\) is then trained on \(\mathcal{D}_{\text{perm}, j, k}\) and its error is computed as:
\begin{equation}
\text{MAE}_{\text{perm}, j, k} = \text{MAE}(M_{\text{perm}, j, k}, \mathcal{D}_{\text{perm}, j, k})
\end{equation}

Finally, the permutation importance (PI) of the search trend for conspiracy theory \(j\) is calculated to evaluate its contribution to the prediction of hate crime occurrences. It is defined as:

\begin{equation}
PI_j = \min \left( \text{MAE}_{\text{perm}, j}^{(1)}, \text{MAE}_{\text{perm}, j}^{(2)}, \text{MAE}_{\text{perm}, j}^{(3)} \right) - \text{MAE}_{\text{original}}
\end{equation} where \(\text{MAE}_{\text{original}}\) represents the MAE of the model when trained and tested on the unaltered dataset. The terms \(\text{MAE}_{\text{perm}, j}^{(1)}, \text{MAE}_{\text{perm}, j}^{(2)}, \text{MAE}_{\text{perm}, j}^{(3)}\) correspond to the MAE values obtained after randomly permuting the search trend data for conspiracy theory \(j\) three separate times and evaluating the model's performance. The function \(\min(\cdot)\) selects the lowest MAE value from the three permutations, ensuring that the best-case scenario is used to assess the theory’s importance.

A feature is considered significant if \(PI_j > 0\), indicating that the model relies on the temporal structure of the feature for accurate predictions.

For models incorporating lagged features (e.g., \(x_{t-\Delta, j}\) for \(\Delta \in \{-1, 0, 1, 2, 3\}\)), each lagged variant is treated as a distinct feature in the permutation test. For instance, if the lag -1 feature \(x_{t-1, j}\) is permuted, the resulting shuffled dataset is given by:
\begin{equation}
\mathcal{D}_{\text{perm}, j, \Delta, k} = \{(\dots, x_{\pi_k(t)-\Delta, j}, \dots), y_t\}_{t=1}^{T}
\end{equation}
This procedure isolates the predictive contribution of individual temporal lags.

To compare with the baseline, let \(M_{\text{baseline}}\) be a model trained without any conspiracy-theory features. A feature \(j\) is validated if:
\begin{equation}
\text{MAE}_{\text{original}} < \text{MAE}_{\text{baseline}}
\end{equation}
(indicating improved accuracy), and
\begin{equation}
PI_j > 0
\end{equation}
(suggesting that the predictive power depends on the feature’s temporal structure rather than on static numerical properties).

The feature permutation test provides an advantage over conventional validation techniques, such as 10-fold cross-validation, which is commonly used to estimate generalization error in machine learning. Cross-validation assumes that data points are independent and identically distributed, an assumption that does not hold for time-series data \cite{bergmeir2012crossvalidation, del2024robust}. Applying random cross-validation to time-series models can distort temporal dependencies and lead to misleading performance estimates. While alternative validation techniques, such as blocked or rolling-window cross-validation, attempt to mitigate this issue, they do not directly assess whether a specific feature contributes causally to predictions \cite{del2024robust}.

Unlike these approaches, the feature permutation test isolates the predictive impact of individual features while preserving the dataset’s temporal integrity. By disrupting only the time-series order of conspiracy-theory search trends while keeping all other data intact, it ensures that observed accuracy gains result from the temporal pattern rather than static numerical properties. This approach is particularly valuable in high-dimensional settings, where traditional causality tests, such as Granger causality, often produce false positives \cite{castro}.

The effectiveness of feature permutation testing has been widely demonstrated in fields where deep learning models often struggle with interpretability, such as biomedical research, financial forecasting, and environmental science \cite{canovas2009permutations, wang2017permutation}. It aligns with best practices in explainable machine learning, reinforcing the role of post hoc interpretability techniques and supporting the integration of causal reasoning into predictive frameworks \cite{lundberg2017unified, scholkopf2021toward}. By maintaining the integrity of temporal sequences and eliminating the risk of spurious correlations, the feature permutation test ensures that detected relationships between conspiracy theory searches and hate crimes are statistically valid. Unlike methods relying on distributional assumptions or post hoc statistical adjustments, permutation testing directly evaluates whether a given search trend holds a meaningful predictive value.

This framework is particularly well suited for time series analysis, where the preservation of sequential dependencies is critical, and conventional validation methods can disrupt them if not carefully adapted \cite{bergmeir2012crossvalidation, canovas2009permutations, del2024robust}. In this study, the final 20\% of the dataset serves as an out-of-sample benchmark, while permutation testing ensures that the observed accuracy gains originate from the conspiracy theory time-series themselves, rather than artifacts of scale or sampling. By adopting this approach, the study enhances model interpretability and ensures that identified relationships reflect actual behavioral patterns linking online information pollution to offline hate crimes.

\section{Results}\label{sec4}

\begin{figure}[h]
    \centering
    \includegraphics[width=0.9\textwidth]{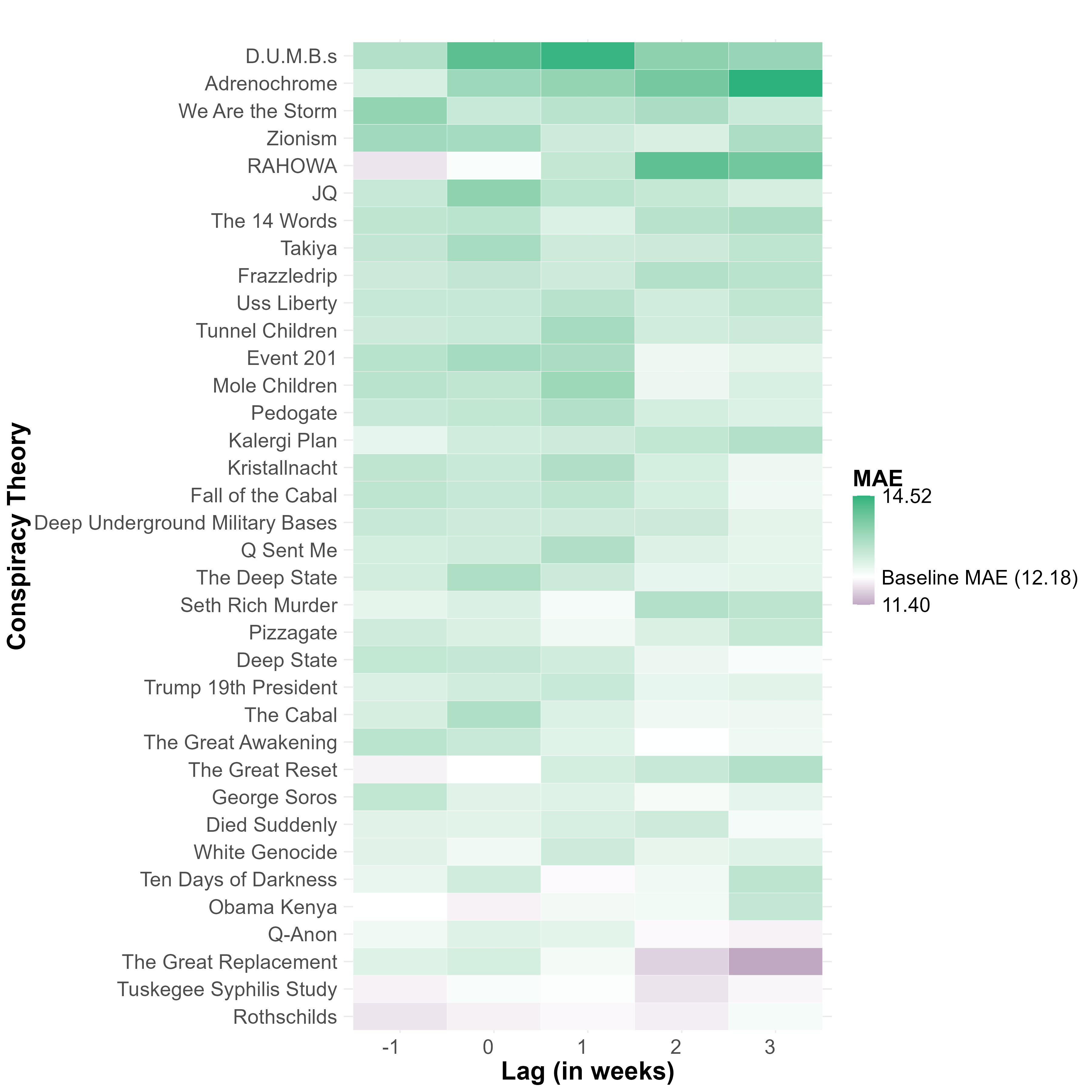}
    \caption{\textbf{Conspiracy Theories' Contributing to Crime Prediction at Various Time Lags} 
    Heat-map of the scaled mean absolute error (MAE) for the 36 conspiracy theories evaluated at various time lags (-1 to 3 weeks). The conspiracy theories are sorted by their average MAE (across lags), from highest on top to lowest on the bottom. Values below the baseline MAE of 12.18—calculated from a model with no conspiracy search time-series—are shaded towards purple, indicating better predictive performance when including information on theories, while values higher than 12.18 are shaded towards green, indicating worse predictive performance.}
    \label{fig4}
\end{figure}

Having tested 36 theories' trends, in 5 different iterations (from lags -1 to 3), we find that for the vast majority of cases, the model does not learn from the trend, that is, the model does not perceive a connection and better predict hate crimes having knowledge of the online demand for 28 conspiracy theories. Among these are popular theories such as \textit{the Kalergi Plan }, \textit{pizzagate}, and \textit{white genocide}. Just as well, search terms that are heavily associated with a range of theories (\textit{George Soros}, \textit{the deep state}, \textit{adrenochrome}) fail to inform the model. That being said, eight theories do register as being able to improve their respective models' predication capacity. All scaled MAEs of the theories tested at each iteration are reported in Figure \ref{fig4}. The conspiracy theories are sorted by their average MAE (across lags), from highest on top to lowest on the bottom. Values below the baseline MAE of 12.18—calculated from a model without conspiracy search time series—are colored purple, indicating better predictive performance when including information on theories, while values higher than 12.18 are colored green, indicating worse predictive performance. 

\FloatBarrier

\textit{The Great Replacement} theory generates the most significant improvement in crime prediction, with reductions in prediction error of 3.29\% and 6.42\% at lags of two and three weeks, respectively (Table \ref{tab2}). The \textit{Rothschilds} theory also displays consistent improvement in prediction across multiple lags, particularly interesting is its 0.99\% decrease in error at lag zero and 2\% decrease when pushing the conspiracy trend forward, This suggests a possible immediate impact or, potentially, the inverse relationship, wherein hate crimes spur greater interest in the conspiracy theory. Similar signals of said inverse relationship were found with the \textit{Tuskegee syphilis study} and \textit{Obama Kenya} both showing a prediction improvement with trends shifting forward. Just as well, the \textit{Tuskegee syphilis study} and \textit{Q-Anon} theories primarily influence the accuracy of the prediction at lags of two and three weeks, with reductions in error of up to 1.89\% and 0.90\%, respectively. The inclusion of the time-series of searches for \textit{Obama Kenya} among predictors comes with a 0.89\% increase in precision. The \textit{Ten days of darkness} theory provides a more modest improvement of 0.33\% solely in lag one. \textit{RAHOWA} and \textit{the Great Reset}, interestingly, improve model accuracy when their respective conspiracy trends are shifted forward, however, not in any other iteration. 
 
\begin{table}[]
\caption{\textbf{Improvements in Prediction}}\label{tab2}
\begin{tabular}{p{4cm}ccccc}
\toprule
\multicolumn{1}{c}{\textbf{Conspiracy Theory}} & \multicolumn{5}{c}{\textbf{Scaled MAE per Lag}} \\
\multicolumn{1}{c}{} & \multicolumn{1}{c}{-1} & \multicolumn{1}{c}{0} & \multicolumn{1}{c}{1} & \multicolumn{1}{c}{2} & \multicolumn{1}{c}{3} \\ \hline
Ten Days of Darkness & 12.45 & 12.78 & 12.13* & 12.37 & 13.01 \\
Obama Kenya & 12.18  & 12.07* & 12.33 & 12.34 & 12.93 \\
Q-Anon & 12.37 & 12.59 & 12.53 & 12.13* & 12.06* \\
The Great Replacement & 12.59 & 12.72 & 12.31 & 11.77* & 11.39* \\
Tuskegee Syphilis Study & 12.06* & 12.27 & 12.22 & 11.94* & 12.09* \\
Rothschilds & 11.95* & 12.05* & 12.12* & 12.03* & 12.30 \\ 
RAHOWA & 11.95* & 12.25 & 12.94 & 14.10 & 13.89 \\ The Great Reset & 12.07* & 12.18 & 12.74 & 12.88 & 13.14 \\\hline
Baseline & \multicolumn{5}{c}{12.18} \\
\hline
\hline
\end{tabular}
\footnotetext{The table displays the scaled mean absolute error (MAE) for impactful conspiracy theories across different time lags (-1 to 3 weeks). It also includes a baseline MAE—calculated from a model with no conspiracy search time—and the result when the lag is reversed (moving the trend for theories ahead of that of crime by one period) for comparison. Values marked with an asterisk (*) indicate that the MAE is lower than the baseline's, suggesting that these conspiracy theories improve the prediction accuracy.}
\end{table}

The permutation test shows that for most conspiracy theories (except \textit{the Great Reset} at lag -1, \textit{Obama Kenya} at lag 0, and \textit{Q-Anon} at lag 2), shuffling the data leads to a worse prediction accuracy (higher MAE) compared to using the original data sequence (Table \ref{tab3}). This suggests that the true data sequences that contain these conspiracy theories actually have predictive power for hate crimes. Therefore, the presence/absence of these conspiracy theories in online discussions seems to be associated with hate crimes.

\begin{table}[]
\centering
\caption{\textbf{Feature Permutation Test}}\label{tab3}
\begin{tabular}{c c c c c c}
\hline
\multicolumn{1}{c}{\textbf{Conspiracy Theory}} & \multicolumn{5}{c}{\textbf{Lowest Scaled MAE from Permuted Runs, per Lag}} \\
& -1 & 0 & 1 & 2 & 3 \\ 
\hline
Ten Days of Darkness  &  &  & 12.73  &  &  \\
                      &  &  & (12.13*) &  &  \\
Obama Kenya           &  &  & 11.72 &  &  \\
                      &  &  & (12.07) &  &  \\
Q-Anon                &  &  &  & 11.96    & 12.24  \\
                      &  &  &  & (12.13)  & (12.06*) \\
The Great Replacement &  &  &  & 12.57    & 12.72  \\
                      &  &  &  & (11.77*) & (11.39*) \\
Tuskegee Syphilis Study & 11.99   &  &  & 13.04    & 12.49  \\
                        & (12.06) &  &  & (11.94*) & (12.09*) \\
Rothschilds            & 12.54    & 12.54    & 12.26    & 12.14    & 12.23  \\
                       & (11.95*) & (11.95*) & (12.05*) & (12.12*) & (12.03*) \\
RAHOWA                 & 12.07  &  &  &  &  \\
                       & (11.95*) &  &  &  &  \\
The Great Reset        & 12.03  &  &  &  &  \\
                       & (12.07) &  &  &  &  \\
\hline
\end{tabular}

\footnotetext{The table presents the results of a feature permutation test, highlighting the lowest scaled mean absolute error (MAE) obtained from permuted runs for impactful conspiracy theories across different time lags (0 to 3 weeks). The values in parentheses represent the MAE with the true data sequence, providing a benchmark for comparison. If the MAE with the true data sequence is lower than the MAE with the permuted data sequence, it suggests that the contribution of the conspiracy theory to the prediction is unlikely to be by chance and it is marked with an asterisk (*).}
\end{table}
\FloatBarrier

\section{Discussion}\label{sec5}

Previous literature suggests that conspiracy theories may create a social environment conducive to extremism and violence \cite{lee, krouwel_does_2017}. In this context, conspiracies may act as radicalizing catalysts, amplifying extremist views and linking them to justifications for violence. People who are receptive to conspiracy beliefs are more likely to endorse violence \cite{jolley_belief_2019, krouwel_does_2017, vegetti2022belief}. This reinforces a societal \textit{us vs. them} mentality, fosters cognitive radicalization through mechanisms such as moral disengagement \cite{frissen}, and encourages behaviors that challenge social norms while aligning with conspiracy narratives. Research indicates that a conspiracy mentality is associated with a decrease in participation in conventional political processes, but an increase in unconventional political actions \cite{imhoff_2020}. Furthermore, exposure to conspiracy theories can intensify feelings of anomie, leading to increased intentions to commit everyday crimes \cite{jolley_exposure_2020}.

Building on this knowledge, our study investigates the connection between online demand for conspiracy theories and the occurrence of offline hate crimes and finds partial empirical support for a link between the conspiracy theories and harmful offline social behaviors, as posited by \cite{bilewicz_harmful_2013}, \cite{krouwel_does_2017}, \cite{lavorgna2022science}, \cite{piazza2022fake}, \cite{wiedlitzka2023hate}. Out of the 36 theories analyzed, eight provide a meaningful improvement in the model's capacity to predict registered hate crime occurrences. Therefore, our expectation is supported--the demand trends for (some) conspiracy theories online can improve the deep learning model’s ability to predict hate crimes offline--but the witnessed relationship is complex, topic dependent, and there is no clear causal direction. 

Therefore, our empirical analysis shows that most conspiracy theories have no clear link to offline hate crimes, but a specific few do show a significant connection. The non-uniformity of prediction improvement suggests that only certain narratives or specific communication styles are associated with changes in real-world behaviors. Some theories may resonate more deeply with individuals predisposed to commit hate crimes, while others may circulate without showing any significant association with offline actions, similar to the observed behavior of hate-speech \cite[e.g.,][]{arcila2024online, williams_hate_2019}.

In terms of magnitude, the observed prediction improvements are modest, but noteworthy. The reduction in MAE-a measure of the average difference between predicted and actual values-indicates improved predictive accuracy for models incorporating specific conspiracy theories compared to the baseline model. These improvements in MAE ranged from 0.33\% to 6.42\%. Although percentages may appear small, they hold importance given the complexity and multitude of factors related to hate crime occurrences \cite[see][]{berkowitz2005hate, king2013high}.

We observe that trends for conspiracies with clear racial narratives provide the most significant improvements, such as the \textit{Rothschilds} which better informed the model in four of the five iterations. \textit{Rothschilds} theory falsely accuses the Rothschild family of controlling global finances and manipulating global events and sovereign nations. This result comes as no surprise as racism and xenophobia are root causes of hate crimes, particularly in the context of far-right ideologies \cite{adamczyk2014relationship, bowes1990racism, lantz2023anti}.

Conversely, \textit{Q-Anon} is the only theory that improves predictions without explicitly focusing on race or minorities, with figures like Celine Dion and Oprah Winfrey being equally implicated.\textit{Ten Days of Darkness}, while potentially non-racially focused and heavily tied into the Q-Anon movement, promises the arrival of a civil war-like conflict and the 'taking of revenge' on a mostly (though not exclusively) Jewish set of elites. That being said, whether these distinct effects are due to the content, timing, spread modalities, or intensity of conspiracy theories is beyond the scope of this study and warrants further research. 

Just as well, our analysis also reveals that some conspiracy theories are associated with delayed patterns in hate crimes, as improvements in predictive accuracy were observed at lags of two to three weeks. In particular, \textit{the Great Replacement}, the belief that Western elites are organizing to replace their white populations with non-white immigrants, showed relatively substantial prediction improvement capacity at longer lags. This finding may indicate that the relationship between these narratives and hate crimes becomes more pronounced over time, though it remains to be clarified whether this is due to the fact that the influence of these narratives builds over time.

In fact, research on the nexus between hate-speech and hate crime struggled to identify the direction of the relationship, believing that the commission of hate crimes offline could easily be interpreted as driver of online hate-speech rather than the reverse, expected direction for the relationship \cite{williams_hate_2019, williams_cyberhate_2016}. In our study, there are similar signals of a reverse relationship. \textit{Rothschilds} and \textit{Tuskegee syphilis study}, as well as \textit{RAHOWA} and \textit{the Great Reset} (which failed to improve accuracy in any of the lags), improve the model when the trends are pushed forward in time - though at said lag only the \textit{Rothschilds} and \textit{RAHOWA} prove valid after feature permutation testing. 

As will be touched on in the proceeding limitations section, the identified relationship can not be stated as causal. There is the potential that an unexamined variable is driving the movement of both studied phenomena. Nevertheless, the aforementioned studies on conspiracy theories, as well as theoretical explanations for criminal behavior, provide a reason to believe that the relationship indicated in our results are per se meaningful. 

\cite{sykes1957techniques} and their \textit{neutralization theory} argue that deviant behavior can emerge when individuals rationalize or neutralize their actions within a system of justifications, effectively granting themselves permission to bypass societal norms. Conspiracy theories may provide such justifications, offering intricate narratives that portray members of society or its institutions as guilty of violating societal norms. For example, \cite{jolley_belief_2019} specifically suggested that believers in conspiracies might start to view the world through a conspiratorial lens and engage in actions they see as appropriate for such a reality. This perspective could erode social barriers to deviant acts, particularly if the authorities enforcing norms are perceived to be committing even greater violations themselves.

Theories like \textit{the Great Replacement} further illustrate this dynamic by expanding the range of responses a prejudiced individual might consider against an out-group. For instance, an individual with discriminatory beliefs about immigrant communities (e.g., stereotyping them as lazy or disruptive neighbors) might struggle to justify hateful actions. However, if immigrants are framed as an existential threat to the survival of the in-group, as this theory suggests, justifying harmful actions becomes far easier.

These neutralization mechanisms may operate regardless of whether the search for conspiracy theories occurs before or after the offline criminal act, highlighting their role in shaping perceptions and justifications rather than serving as direct causal triggers.

Continuing with in-group/out-group dynamics, \textit{differential association} theory posits that criminal behavior is learned through interactions with others, during which values, attitudes, and rationalizations that support criminal acts are transferred \cite{sutherland1947principles}. The spread of conspiracy theories within ideologically homogeneous online and offline communities may act as a mechanism for such transmission. Studies on the spread of medical misinformation highlight the role of friendship networks among misinformation consumers, characterized by frequent interactions and strong bonds within groups \cite{lavorgna2022science, tweetandquack}.

The formation of an in-group, those who perceive themselves as possessing the \textit{hidden truth}, can create an environment where hate crimes against the out-group are rationalized and deemed acceptable. We might think that conspiracy theories themselves may act as the transmitters of a differential set of values, attitudes, and justifications, reshaping the worldview of a new adopter in ways that reduce barriers to deviant behavior, even in the absence of direct interpersonal interaction.

\subsection{Limitations and Future Research}\label{5.2}

Our study has several limitations that should be considered when interpreting our results. First, external confounding factors were not considered. We show that the active searching for certain conspiracy theories is related to actual offline episodes of hate and that peaks in searches for specific conspiracies precede increases in hate crimes. Relying on theory and previous findings, we argue that it is plausible that conspiracies foster anomie and, thus, create the ideological conditions for extremism and violence. However, our analysis does not rule out the possibility that the demand for conspiracy theories is a mediating factor between other unseen factors and hate crimes. Individual psychosocial characteristics, as well as the general political climate, might influence both the demand for conspiracies and the commission of hate crimes. Investigating the psychological and sociopolitical factors that mediate this relationship could provide valuable insights and should attract further research.

Secondly, using searches for keywords on Google to proxy the online demand for conspiracies is imperfect. Searches for certain keywords might be driven both by the search for conspiracies related to them and by genuine searches for the same topic irrespective of the connected conspiracy. For example, the Tuskegee syphilis study is a real historical event and a dark chapter in American history. Among conspiracy circles, it is often used as justification/evidence in favor of conspiratorial beliefs regarding the American healthcare system \cite[]{regina}. While it is plausible to assume that nowadays the majority of searches are motivated by an interest in the conspiracy theory, we can not rule out the possibility of searches for the actual event. Similarly, George Soros and the Rothschild family are protagonists in numerous theories, though part of the Google searches for the keywords \textit{George Soros} and \textit{Rothschilds} are likely not related to any conspiracy but rather to the persons themselves. The dual nature of these search prompts might introduce a downward bias in the estimate of the impact of conspiracies on prediction; the more searches unrelated to the conspiracy, the more severe the downward bias.

Another limitation in the use of Google Trends data is the possible use of alternative keywords to signify the same theory. For instance, both \textit{seth rich murder} and \textit{seth rich} are commonly used labels to refer to the \textit{Seth Rich murder} conspiracy theory, which posits that Seth Rich leaked Democratic National Committee emails to WikiLeaks and was killed for it, not shot during a robbery attempt, as claimed by law enforcement authorities. Similarly, \textit{takiya}, \textit{taqiya}, and \textit{taqiyya} might all be used to refer to a conspiracy theory that misrepresents a concept in Islam (i.e., dissimulation to protect oneself) wherein Muslims can lie to deceive non-Muslims and advance the spread of the religion. We checked for the alternative use of these keywords, and our results did not change. However, it might be that while the distinct time-series referring to these keywords are not associated with variations in hate crime counts, the combination of searches for these alternative terms does relate to the number of registered hate crimes. In general, our analysis does not assess the possible simultaneous impact of multiple conspiracies. We adopted this approach to allow for differentiated results by specific theory, thus accounting for their heterogeneity. Future research can add to our own by analyzing the potential simultaneous impact of multiple theories. Similarly, scholars might evaluate information pollution topics more associated with \textit{fake news}, as opposed to conspiracy theories.

A further limitation of our approach lies in its sole reliance on Google Trends to measure online interest in conspiracy theories. Prior research underscores the role of active information-seeking in radicalization processes \cite{frissen} and suggests that Google Trends is a coherent tool for examining broad online interest \cite{ahmed2017financial, behnert2024can, bong2020analysis, prado2021google, mager2023european, morsy2018prediction}. Nonetheless, this method does not account for searches conducted on social media platforms (e.g., Instagram, Twitter, TikTok, Reddit). These platforms are particularly popular among younger demographics, who may exhibit distinct search behaviors and differing levels of engagement with conspiratorial content. Consequently, this exclusion may result in unaccounted variance in online information-seeking. Although searches on social media and Google are likely correlated, they are imperfectly so. Future studies could address this gap by integrating data from multiple platforms to better trace how demand for conspiratorial content translates into offline hate crimes.

Whilst our focus was on finding the presence of a relationship, researchers could subsequently examine the connected theories in-depth to understand any overlaps in their content and/or structure. In doing so, it may be possible to identify what contributes to a theory's transference from the online to the offline world. Contributing to this effort, future analysis may choose to examine hate crimes at a more granular level. Depending on the statutory government, hate crime data may be annotated with the offenders bias (i.e., the specific hate behind each offense). Instead of a time-series prediction, or causal discovery approach, a deep learning classification task can be used to help identify if flows in demand for conspiracy theories, with a given bias, help the model to understand when hate crimes with the same bias will occur.

Moving to the crime evaluated, a well-documented issue of official hate crime data is under-reporting, with victims often reluctant to come forward due to fear of retaliation, distrust in law enforcement, or lack of awareness that their experience constitutes a hate crime. This underestimates the true prevalence of hate crimes \cite{pezzella2019dark}. Additionally, unconscious biases held by both victims and law enforcement can introduce inconsistencies in reporting \cite{vergani2023hate}. In studying the relationship between conspiracy theories and hate crimes, these limitations are relevant if significant changes in hate crime reporting practices occurred during the study period. However, we currently do not have any evidence of such substantial shifts in Michigan during the chosen period.

Finally, we conducted our analysis at the state level, focusing on Michigan, which has a population of around 10.1 million and contexts as heterogeneous as Highland Park and Bloomfield Hills. The relationship between conspiracy theories and hate crimes may vary based on factors like the marginalization of minorities and socio-economic disparities. Although our aim was to provide a general overview, the research field would benefit from more localized studies to understand these geographical variations. Moreover, exploring the impact of conspiracy theories in different geographical and cultural contexts would enhance our understanding of their global implications and provide insight into the role of socio-economic confounding variables, which, due to their annual reporting, are not well-suited for a deep learning model. Observing the results in different geo-political regions with varying socio-economic conditions would allow for comparisons between common and divergent circumstances, contributing to an understanding of whether socio-economic conditions play a confounding role in the transmission of conspiracy theories, both on and offline.

Related to this, it is important to note that our reliance on Google Trends' IP-based geolocation introduces the possibility of geographic misclassification. Users employing VPNs or temporarily located outside Michigan but maintaining local search behavior could be misattributed. Moreover, IP geolocation accuracy varies depending on network types and connection density, and it can not guarantee precise location at the city or regional level. For example, while specific error rates fluctuate, services like MaxMind's GeoIP2 demonstrate a median error of 2.6 km for fixed-line addresses in New York City \cite{saxon2022gps}. Therefore, our findings may under represent transient populations or overrepresent residents with more stable IP addresses. However, research suggests that geolocation of IP addresses is often reliable for determining the country \cite{komosny2017location}. Thus, IP addresses are generally a robust method for discerning state-level trends despite this potential noise.

\section{Conclusions}\label{sec6}

This study addresses a gap in existing research by exploring the relationship between the online demand for conspiracy theories and the occurrence of offline hate crimes. Previous research has primarily focused on the spread of conspiracy theories and their impact on perceptions and beliefs, but there has been limited empirical investigation into their direct influence on real-world criminal behavior. Moreover, this study shifts the focus from the supply of disinformation (articles, blogs, and posts) to the demand side, providing a novel approach to understanding the potential connection between the online demand for conspiracy theories, captured through search trends, and the commission of offline hate crimes.

Our approach offers a perspective on detecting feature importance by using deep learning beyond pattern recognition. Specifically analyzing the online search demand for 36 racially- and politically-charged conspiracy theories, we forecast hate crime occurrences in Michigan. The results of which reveal that, while most conspiracy theories do not show a significant relationship with hate crimes, a subset of eight, including \textit{The Great Replacement}, \textit{Rothschilds} family, and \textit{Q-Anon}, does improve our models' predictive power.

These findings underscore the need for targeted interventions to address the spread of specific harmful conspiracy theories. Recognizing that general efforts to combat online information pollution are crucial, focusing on the most impactful narratives could be the more effective strategy to prevent offline violence. Policymakers and social media platforms should prioritize monitoring and countering the spread of conspiracy theories that are strongly linked to hate crimes, taking a targeted approach rather than spreading resources thin.

Furthermore, our results indicate an association between the spread of certain conspiracy theories and societal challenges, including patterns of criminal acts. Additionally, our findings demonstrate how deep learning algorithms can be leveraged to provide insight into even the most complex social mechanisms.

Our research also suggests the role of conspiracy theories in fostering ideological conditions conducive to extremism and violence. Forms of prejudice reinforce extremist views and divisive mentalities, potentially leading to real-world criminal behavior. These dynamics align with previous research suggesting that conspiracy beliefs can lead to radicalization and anti-normative actions. Furthermore, the study demonstrates that the relation of conspiracy theories on hate crimes can be delayed, with significant effects on prediction capacity observed weeks after increased online interest.

\section*{Data Availability}
All data used in the research are publicly available and accessible at the URL specified in the \textit{Methods} section.

\section*{Contributing Authors}
The paper is the result of the joint work of Alberto Aziani, Michael V. LoGiudice, and Ali Shadman Yazdi, who together conceived the presented idea. A.A. supervised the findings of this work. M.V.L. managed the data collection and the literature review. A.S.Y. performed the computations. All authors discussed the results. A.A. and M.V.L. contributed to the final manuscript.

\section*{Conflict of Interest}
The authors have no competing financial or non-financial interests to disclose.

\bibliographystyle{apacite}
\bibliography{BIBCTC}

\end{document}